\documentclass[aps,prl,twocolumn,groupedaddress,showpacs,nofootinbib,final,superscriptaddress,preprintnumbers,floatfix]{revtex4}
\usepackage{graphicx}
\usepackage{amssymb}
\usepackage{stmaryrd}
\usepackage{amsmath}
\begin{document}

\newcommand{\lambdal}[0]{\lambda_{\text{L}}}
\newcommand{\omegad}[0]{\omega_{\text{D}}}
\newcommand{\bc}[0]{B_{\text{c}}}
\newcommand{\nbb}[0]{{0\nu}\beta\beta}
\newcommand{\den}[0]{d_{\text{e}}^{\cal{N}}}
\newcommand{\fp}[0]{f_\pi}
\newcommand{\eq}[1]{eq.~(\ref{#1})}
\newcommand{\eqs}[2]{eqs.~(\ref{#1,#2})}
\newcommand{\Eq}[1]{Eq.~(\ref{#1})}
\newcommand{\Eqs}[2]{Eqs.~(\ref{#1},\ref{#2})}
\newcommand{\fig}[1]{fig.~\ref{#1}}
\newcommand{\figs}[2]{figs.~\ref{#1},\ref{#2}}
\newcommand{\Fig}[1]{Fig.~\ref{#1}}
\newcommand{\qbb}[0]{Q_{\beta\beta}}
\newcommand{\tl}[0]{\text{L}}
\newcommand{\tr}[0]{\text{R}}
\newcommand{\nc}[0]{N_{\text{c}}}
\newcommand{\mw}[0]{M_{\text{W}}}
\newcommand{\mz}[0]{M_{\text{Z}}}
\newcommand{\mr}[0]{M_{\text{R}}}
\newcommand{\md}[0]{m_{\text{D}}}
\newcommand{\mn}[0]{m_\nu}
\newcommand{\lh}[0]{\Lambda_{\text{H}}}
\newcommand{\aif}[0]{a^{ff^\prime}_{i;ll^\prime}}
\newcommand{\gs}[0]{\Gamma_{\text{S}}=1}
\newcommand{\gps}[0]{\Gamma_{\text{PS}}=\gamma^5}
\newcommand{\lr}[0]{\Lambda_{\text{R}}}
\newcommand{\wrt}[0]{W_{\text{R}}}
\newcommand{\wl}[0]{W_{\text{L}}}
\newcommand{\ls}[0]{\Lambda_{\text{S}}}
\newcommand{\gf}[0]{G_{\text{F}}}
\newcommand{\mm}[0]{M_{\text{M}}}
\newcommand{\sst}[1]{{\scriptscriptstyle #1}}
\newcommand{\beq}{\begin{equation}}
\newcommand{\eeq}{\end{equation}}
\newcommand{\beqa}{\begin{eqnarray}}
\newcommand{\eeqa}{\end{eqnarray}}
\newcommand{\dida}[1]{/ \!\!\! #1}
\renewcommand{\Im}{\mbox{\sl{Im}}}
\renewcommand{\Re}{\mbox{\sl{Re}}}
\def\simge{\hspace*{0.2em}\raisebox{0.5ex}{$>$}
     \hspace{-0.8em}\raisebox{-0.3em}{$\sim$}\hspace*{0.2em}}
\def\simle{\hspace*{0.2em}\raisebox{0.5ex}{$<$}
     \hspace{-0.8em}\raisebox{-0.3em}{$\sim$}\hspace*{0.2em}}
\def\dn{{d_n}}
\def\de{{d_e}}
\def\datom{{d_{\sst{A}}}}
\def\grhobar{{{\bar g}_\rho}}
\def\gpibar{{{\bar g}_\pi^{(I) \prime}}}
\def\gpibarz{{{\bar g}_\pi^{(0) \prime}}}
\def\gpibaro{{{\bar g}_\pi^{(1) \prime}}}
\def\gpibart{{{\bar g}_\pi^{(2) \prime}}}
\def\mx{{M_X}}
\def\mrho{{m_\rho}}
\def\qpv{{Q_{\sst{W}}}}
\def\lamtv{{\Lambda_{\sst{TVPC}}}}
\def\lamtvs{{\Lambda_{\sst{TVPC}}^2}}
\def\lamtvc{{\Lambda_{\sst{TVPC}}^3}}

\def\bra#1{{\langle#1\vert}}
\def\ket#1{{\vert#1\rangle}}
\def\coeff#1#2{{\scriptstyle{#1\over #2}}}
\def\undertext#1{{$\underline{\hbox{#1}}$}}
\def\hcal#1{{\hbox{\cal #1}}}
\def\sst#1{{\scriptscriptstyle #1}}
\def\eexp#1{{\hbox{e}^{#1}}}
\def\rbra#1{{\langle #1 \vert\!\vert}}
\def\rket#1{{\vert\!\vert #1\rangle}}

\def\lsim{{ <\atop\sim}}
\def\gsim{{ >\atop\sim}}
\def\nubar{{\bar\nu}}
\def\psibar{{\bar\psi}}
\def\Gmu{{G_\mu}}
\def\alr{{A_\sst{LR}}}
\def\wpv{{W^\sst{PV}}}
\def\evec{{\vec e}}
\def\notq{{\not\! q}}
\def\notl{{\not\! \ell}}
\def\notk{{\not\! k}}
\def\notp{{\not\! p}}
\def\notpp{{\not\! p'}}
\def\notder{{\not\! \partial}}
\def\notcder{{\not\!\! D}}
\def\notA{{\not\!\! A}}
\def\notv{{\not\!\! v}}
\def\Jem{{J_\mu^{em}}}
\def\Jana{{J_{\mu 5}^{anapole}}}
\def\nue{{\nu_e}}
\def\mns{{m^2_{\sst{N}}}}
\def\me{{m_e}}
\def\mes{{m^2_e}}
\def\mq{{m_q}}
\def\mqs{{m_q^2}}
\def\mw{{M_{\sst{W}}}}
\def\mz{{M_{\sst{Z}}}}
\def\mzs{{M^2_{\sst{Z}}}}
\def\ubar{{\bar u}}
\def\dbar{{\bar d}}
\def\sbar{{\bar s}}
\def\qbar{{\bar q}}
\def\sstw{{\sin^2\theta_{\sst{W}}}}
\def\gv{{g_{\sst{V}}}}
\def\ga{{g_{\sst{A}}}}
\def\pv{{\vec p}}
\def\pvs{{{\vec p}^{\>2}}}
\def\ppv{{{\vec p}^{\>\prime}}}
\def\ppvs{{{\vec p}^{\>\prime\>2}}}
\def\qv{{\vec q}}
\def\qvs{{{\vec q}^{\>2}}}
\def\xv{{\vec x}}
\def\xpv{{{\vec x}^{\>\prime}}}
\def\yv{{\vec y}}
\def\tauv{{\vec\tau}}
\def\sigv{{\vec\sigma}}

\def\sst#1{{\scriptscriptstyle #1}}
\def\gpnn{{g_{\sst{NN}\pi}}}
\def\grnn{{g_{\sst{NN}\rho}}}
\def\gnnm{{g_{\sst{NNM}}}}
\def\hnnm{{h_{\sst{NNM}}}}
\def\xivz{{\xi_\sst{V}^{(0)}}}
\def\xivt{{\xi_\sst{V}^{(3)}}}
\def\xive{{\xi_\sst{V}^{(8)}}}
\def\xiaz{{\xi_\sst{A}^{(0)}}}
\def\xiat{{\xi_\sst{A}^{(3)}}}
\def\xiae{{\xi_\sst{A}^{(8)}}}
\def\xivtez{{\xi_\sst{V}^{T=0}}}
\def\xivteo{{\xi_\sst{V}^{T=1}}}
\def\xiatez{{\xi_\sst{A}^{T=0}}}
\def\xiateo{{\xi_\sst{A}^{T=1}}}
\def\xiva{{\xi_\sst{V,A}}}
\def\rvz{{R_{\sst{V}}^{(0)}}}
\def\rvt{{R_{\sst{V}}^{(3)}}}
\def\rve{{R_{\sst{V}}^{(8)}}}
\def\raz{{R_{\sst{A}}^{(0)}}}
\def\rat{{R_{\sst{A}}^{(3)}}}
\def\rae{{R_{\sst{A}}^{(8)}}}
\def\rvtez{{R_{\sst{V}}^{T=0}}}
\def\rvteo{{R_{\sst{V}}^{T=1}}}
\def\ratez{{R_{\sst{A}}^{T=0}}}
\def\rateo{{R_{\sst{A}}^{T=1}}}
\def\mro{{m_\rho}}
\def\mks{{m_{\sst{K}}^2}}
\def\mpi{{m_\pi}}
\def\mpis{{m_\pi^2}}
\def\mom{{m_\omega}}
\def\mphi{{m_\phi}}
\def\Qhat{{\hat Q}}
\def\FOS{{F_1^{(s)}}}
\def\FTS{{F_2^{(s)}}}
\def\GAS{{G_{\sst{A}}^{(s)}}}
\def\GES{{G_{\sst{E}}^{(s)}}}
\def\GMS{{G_{\sst{M}}^{(s)}}}
\def\GATEZ{{G_{\sst{A}}^{\sst{T}=0}}}
\def\GATEO{{G_{\sst{A}}^{\sst{T}=1}}}
\def\mdax{{M_{\sst{A}}}}
\def\mustr{{\mu_s}}
\def\rsstr{{r^2_s}}
\def\rhostr{{\rho_s}}
\def\GEG{{G_{\sst{E}}^\gamma}}
\def\GEZ{{G_{\sst{E}}^\sst{Z}}}
\def\GMG{{G_{\sst{M}}^\gamma}}
\def\GMZ{{G_{\sst{M}}^\sst{Z}}}
\def\GEn{{G_{\sst{E}}^n}}
\def\GEp{{G_{\sst{E}}^p}}
\def\GMn{{G_{\sst{M}}^n}}
\def\GMp{{G_{\sst{M}}^p}}
\def\GAp{{G_{\sst{A}}^p}}
\def\GAn{{G_{\sst{A}}^n}}
\def\GA{{G_{\sst{A}}}}
\def\GETEZ{{G_{\sst{E}}^{\sst{T}=0}}}
\def\GETEO{{G_{\sst{E}}^{\sst{T}=1}}}
\def\GMTEZ{{G_{\sst{M}}^{\sst{T}=0}}}
\def\GMTEO{{G_{\sst{M}}^{\sst{T}=1}}}
\def\lamd{{\lambda_{\sst{D}}^\sst{V}}}
\def\lamn{{\lambda_n}}
\def\lams{{\lambda_{\sst{E}}^{(s)}}}
\def\bvz{{\beta_{\sst{V}}^0}}
\def\bvo{{\beta_{\sst{V}}^1}}
\def\Gdip{{G_{\sst{D}}^\sst{V}}}
\def\GdipA{{G_{\sst{D}}^\sst{A}}}
\def\fks{{F_{\sst{K}}^{(s)}}}
\def\FIS{{F_i^{(s)}}}
\def\fpi{{F_\pi}}
\def\fk{{F_{\sst{K}}}}
\def\RAp{{R_{\sst{A}}^p}}
\def\RAn{{R_{\sst{A}}^n}}
\def\RVp{{R_{\sst{V}}^p}}
\def\RVn{{R_{\sst{V}}^n}}
\def\rva{{R_{\sst{V,A}}}}
\def\xbb{{x_B}}
\def\mlq{{M_{\sst{LQ}}}}
\def\mlqs{{M_{\sst{LQ}}^2}}
\def\lscal{{\lambda_{\sst{S}}}}
\def\lvect{{\lambda_{\sst{V}}}}
\def\PR#1{{{\em   Phys. Rev.} {\bf #1} }}
\def\PRC#1{{{\em   Phys. Rev.} {\bf C#1} }}
\def\PRD#1{{{\em   Phys. Rev.} {\bf D#1} }}
\def\PRL#1{{{\em   Phys. Rev. Lett.} {\bf #1} }}
\def\NPA#1{{{\em   Nucl. Phys.} {\bf A#1} }}
\def\NPB#1{{{\em   Nucl. Phys.} {\bf B#1} }}
\def\AoP#1{{{\em   Ann. of Phys.} {\bf #1} }}
\def\PRp#1{{{\em   Phys. Reports} {\bf #1} }}
\def\PLB#1{{{\em   Phys. Lett.} {\bf B#1} }}
\def\ZPA#1{{{\em   Z. f\"ur Phys.} {\bf A#1} }}
\def\ZPC#1{{{\em   Z. f\"ur Phys.} {\bf C#1} }}
\def\etal{{{\em   et al.}}}
\def\delalr{{{delta\alr\over\alr}}}
\def\pbar{{\bar{p}}}
\def\lamchi{{\Lambda_\chi}}
\def\qw0{{Q_{\sst{W}}^0}}
\def\qwp{{Q_{\sst{W}}^P}}
\def\qwn{{Q_{\sst{W}}^N}}
\def\qwe{{Q_{\sst{W}}^e}}
\def\qem{{Q_{\sst{EM}}}}
\def\gae{{g_{\sst{A}}^e}}
\def\gve{{g_{\sst{V}}^e}}
\def\gvf{{g_{\sst{V}}^f}}
\def\gaf{{g_{\sst{A}}^f}}
\def\gvu{{g_{\sst{V}}^u}}
\def\gau{{g_{\sst{A}}^u}}
\def\gvd{{g_{\sst{V}}^d}}
\def\gad{{g_{\sst{A}}^d}}
\def\gvftil{{\tilde g_{\sst{V}}^f}}
\def\gaftil{{\tilde g_{\sst{A}}^f}}
\def\gvetil{{\tilde g_{\sst{V}}^e}}
\def\gaetil{{\tilde g_{\sst{A}}^e}}
\def\gvqtil{{\tilde g_{\sst{V}}^e}}
\def\gaqtil{{\tilde g_{\sst{A}}^e}}
\def\gvutil{{\tilde g_{\sst{V}}^e}}
\def\gautil{{\tilde g_{\sst{A}}^e}}
\def\gvdtil{{\tilde g_{\sst{V}}^e}}
\def\gadtil{{\tilde g_{\sst{A}}^e}}
\def\delp{{\delta_P}}
\def\delzp{{\delta_{00}}}
\def\deld{{\delta_\Delta}}
\def\dele{{\delta_e}}
\def\lnew{{{\cal L}_{\sst{NEW}}}}
\def\osffp{{{\cal O}_{7a}^{ff'}}}
\def\oszg{{{\cal O}_{7c}^{Z\gamma}}}
\def\osgg{{{\cal O}_{7b}^{g\gamma}}}


\def\slash#1{#1\!\!\!{/}}
\def\beq{\begin{eqnarray}}
\def\eeq{\end{eqnarray}}
\def\bea{\begin{eqnarray*}}
\def\eea{\end{eqnarray*}}
\def\NCA{\em Nuovo~Cimento}
\def\IJMP{\em Intl.~J.~Mod.~Phys.}
\def\NP{\em Nucl.~Phys.}
\def\PLB{{\em Phys.~Lett.}~B}
\def\JETPLett{{\em JETP Lett.}}
\def\PRL{\em Phys.~Rev.~Lett.}
\def\MPL{\em Mod.~Phys.~Lett.}
\def\PRD{{\em Phys.~Rev.}~D}
\def\PR{\em Phys.~Rev.}
\def\PRP{\em Phys.~Rep.}
\def\ZPC{{\em Z.~Phys.}~C}
\def\PTP{{\em Prog.~Theor.~Phys.}}
\def\Baryon{{\rm B}}
\def\Lepton{{\rm L}}
\def\sbar{\overline}
\def\stilde{\widetilde}
\def\st{\scriptstyle}
\def\sst{\scriptscriptstyle}
\def\vac{|0\rangle}
\def\argh{{{\rm arg}}}
\def\G{\stilde G}
\def\Wmess{W_{\rm mess}}
\def\NI{\stilde N_1}
\def\antivac{\langle 0|}
\def\infinity{\infty}
\def\mco{\multicolumn}
\def\epp{\epsilon^{\prime}}
\def\psibar{\overline\psi}
\def\nmess{N_5}
\def\chibar{\overline\chi}
\def\lagr{{\cal L}}
\def\drbar{\overline{\rm DR}}
\def\msbar{\overline{\rm MS}}
\def\conj{{{\rm c.c.}}}
\def\Et{{\slashchar{E}_T}}
\def\Etot{{\slashchar{E}}}
\def\mZ{m_Z}
\def\MPlanck{M_{\rm P}}
\def\mW{m_W}
\def\cbeta{c_{\beta}}
\def\sbeta{s_{\beta}}
\def\cW{c_{W}}
\def\sW{s_{W}}
\def\deltaeps{\delta}
\def\sigmabar{\overline\sigma}
\def\epsilonbar{\overline\epsilon}
\def\vep{\varepsilon}
\def\ra{\rightarrow}
\def\half{{1\over 2}}
\def\ko{K^0}
\def\be{\beq}
\def\ee{\eeq}
\def\bea{\begin{eqnarray}}
\def\eea{\end{eqnarray}}
\def\alr{A_{\sst{LR}}}

\def\centeron#1#2{{\setbox0=\hbox{#1}\setbox1=\hbox{#2}\ifdim
\wd1>\wd0\kern.5\wd1\kern-.5\wd0\fi
\copy0\kern-.5\wd0\kern-.5\wd1\copy1\ifdim\wd0>\wd1
\kern.5\wd0\kern-.5\wd1\fi}}
\def\ltap{\;\centeron{\raise.35ex\hbox{$<$}}{\lower.65ex\hbox{$\sim$}}\;}
\def\gtap{\;\centeron{\raise.35ex\hbox{$>$}}{\lower.65ex\hbox{$\sim$}}\;}
\def\gsim{\mathrel{\gtap}}
\def\lsim{\mathrel{\ltap}}
\def\slashchar#1{\setbox0=\hbox{$#1$}           
   \dimen0=\wd0                                 
   \setbox1=\hbox{/} \dimen1=\wd1               
   \ifdim\dimen0>\dimen1                        
      \rlap{\hbox to \dimen0{\hfil/\hfil}}      
      #1                                        
   \else                                        
      \rlap{\hbox to \dimen1{\hfil$#1$\hfil}}   
      /                                         
   \fi}                                        %

\setcounter{tocdepth}{2}







{
\title{A dipole amplifier for electric dipole moments, axion-like particles and a dense dark matter hairs detector} 

\author{Gary Pr{\'e}zeau}\affiliation{Jet Propulsion
  Laboratory, California Institute of  Technology, 4800 Oak Grove Dr,
  Pasadena, CA 91109, USA}


\begin{abstract}
A tool that can constrain, in minutes, beyond-the-standard-model parameters like electric dipole moments (EDM) down to a lower-bound $\den<10^{-37}\text{e}\cdot\text{cm}$ in bulk materials, or the coupling of axion-like particles (ALP) to photons down to $|G_{a\gamma\gamma}|<10^{-16}$~GeV$^{-1}$, is described.  Best limits are $d^n_e<3\cdot10^{-26}\text{e}\cdot\text{cm}$ for neutron EDM and $|G_{a\gamma\gamma}|<6.6\cdot10^{-11}$~GeV$^{-1}$.  The {\it dipole amplifier} is built from a superconducting loop immersed in a toroidal magnetic field, $\vec{B}$.  When nuclear magnetic moments in the London penetration depth align with $\vec{B}$, the bulk magnetization is always accompanied by an EDM-induced bulk electric field $\vec{E}\propto\vec{B}$ that generates detectable oscillatory supercurrents with a characteristic frequency $\omega_{\text{D}}\propto\den$. Cold dark matter (CDM) ALP are formally similar where $\omegad\propto |G_{a\gamma\gamma}|\sqrt{n_a/(2m_a)}$ with $m_a$ the ALP mass and $n_a$ its number density.  A space probe traversing a dark matter hair with a dipole amplifier is sensitive enough to detect ALP density variations if $|G_{a\gamma\gamma}|\sqrt{n_h/(2m_a)}>4.9\cdot10^{-27}$ where $n_h$ is the ALP number density in the hair.
\end{abstract}


\maketitle
}

Experimental searches for the axion and the neutron EDM have been ongoing for decades, motivated by what they could reveal about physics beyond the standard model~\cite{Peccei:1977ur,Pospelov:2005pr}.  Indeed, the axion is a particle thought to have been created when the universe was as young as $10^{-29}$~s; it was originally proposed to explain the {\it strong CP problem}, by which the neutron EDM appears to be infinitesimal despite the CP-violating terms in the QCD Lagrangian.  Its  link to QCD and its appearance early in the universe made the axion a popular CDM candidate.

Consider an effective Lagrangian for CDM ALP, an external electromagnetic field, and an atom in a bulk material that includes P-odd/T-odd~($\slash{P}\slash{T}$) interactions
\beq\label{efflag}
\mathcal{L}_{\text{eff}}&=& { \cal{L}_{\text{K.E.}} }  - \frac{1}{4\mu_0}F_{\mu\nu}F^{\mu\nu} - A_\mu J^\mu_{\text{free}} + \frac{1}{2}F_{\mu\nu}M^{\mu\nu}  \nonumber \\
& &  +\left[ \text{V}_\text{ES} + \text{V}_{Ne} + \text{V}_{NN}+ \frac{d^p_{\text{e}}}{2}\frac{\bar{\psi}_p i\sigma^{\mu\nu}\gamma_5\psi_p}{\mu_p} cF_{\mu\nu} \right] \nonumber \\
& & -\frac{G_{a\gamma\gamma}^i a_i}{4}\frac{F_{\mu\nu}\tilde{F}^{\mu\nu}}{\mu_0} ~,
\eeq
where  $\cal{L}_\text{K.E.}$ contains the kinetic energy terms,  $\text{V}_\text{ES}$ is the atomic electrostatic potential energy, $\text{V}_{Ne}$ has $\slash{P}\slash{T}$ nucleon-electron interactions, $\text{V}_{NN}$ represents the nuclear potential energy including $\slash{P}\slash{T}$ $NN$ interactions, $a_i$ is an ALP field, $G_{a\gamma\gamma}^i$ is an ALP coupling constant to photons,  $d^p_{\text{e}}$ is the electric dipole moment of field $p=\text{e,p,n}$, $\psi_p$ is the wave function of the field $p$, and $\mu_p$ its magnetic moment;  the sums over $i$ and $p$ are  implicit. $A_\mu$ is the photon field, $J^\mu_{\text{free}}$ represents the free current density, $\tilde{F}^{\mu\nu}=\epsilon^{\alpha\beta\mu\nu}F_{\alpha\beta}$ is the dual electromagnetic tensor, and the material's magnetization-polarization tensor (MPT) is given by $M^{0i}=cP^i$, $M^{ij}=-\epsilon^{ijk}M^k$ and $M^{\mu\nu}=-M^{\nu\mu}$ with $\vec{P}$ the bulk electric polarization and $\vec{M}$ the bulk magnetization of the material.

The $\slash{P}\slash{T}$ interactions manifest themselves differently depending on whether an atom is paramagnetic or diamagnetic.  Following the arguments in Refs.~\cite{Spevak:1996tu} in the case of a diamagnetic atom like $^{199}$Hg, and neglecting the electron EDM, the multipole expansion of $V_\text{ES}$ leads to a cancellation of the EDM terms in the brackets to leading-order.  The non-zero terms are higher multipoles such as the Schiff moments.  The $\slash{P}\slash{T}$ moments  are $\propto \mu_\text{s}{d_\text{e}^{\cal{N}}} \vec{I}$ where $\vec{I}$ is the angular momentum of the nucleus, $d_\text{e}^{\cal{N}}$ is the nuclear EDM suppressed by $\mu_\text{s}$, a small dimensionless factor representing the screening of  $d_\text{e}^{\cal{N}}$ by the electron cloud.  Depending on the atom, Schiff moments can be very difficult to calculate and we parameterize the EDM suppression with the order-of-magnitude formula $\mu_{\text{s}}\sim 10Z^2 r_{\text{atom}}^2/r_{\text{nuc}}^2$ where $r_{\text{atom}}$ is the atomic radius, $r_{\text{nuc}}$ is the nuclear charge radius and $Z$ the atomic number; this is conservative since it doesn't account for potential octopole enhancements.  Screening differences between vaporized atoms and solids are not  considered.  The bracketed $\slash{P}\slash{T}$ contributions of the second line in \Eq{efflag} can be replaced by
\beq
\sum_{\alpha=\text{ES},Ne,NN}\!\!\!\!\!\!\!\!\!\!\!\text{V}_\alpha\!+\! \frac{d^p_{\text{e}}\!\bar{\psi}_p i\sigma^{\mu\nu}\!\gamma_5\psi_p}{2\mu_p} cF_{\mu\nu} \!\doteq\! \frac{\mu_\text{s} d_\text{e}^{\cal{N}}\! \bar{\psi} i\sigma^{\mu\nu}\!\gamma_5\psi}{2 \mu_{\cal{N}}} cF_{\mu\nu}
\eeq
where $\psi$ is the nuclear wave function and $\mu_{\cal{N}}$ is the nuclear magnetic moment. Keeping only the ALP + $\slash{P}\slash{T}$ + electromagnetic terms, \Eq{efflag} becomes
\beq\label{effEM}
\mathcal{L}_{\text{eff}} &\doteq&  - \frac{1}{4\mu_0}F_{\mu\nu}F^{\mu\nu} - A_\mu J^\mu_{\text{free}} + \frac{1}{2}F_{\mu\nu}\tilde{M}^{\mu\nu}~~~~~ \\ \label{MPT}
\tilde{M}^{\mu\nu} &=& M^{\mu\nu}   -\frac{G_{a\gamma\gamma}^i a_i}{2}\frac{\tilde{F}^{\mu\nu}}{\mu_0} + c\frac{\mu_\text{s}  d_\text{e}^{\cal{N}}\! \bar{\psi} i\sigma^{\mu\nu}\!\gamma_5\psi}{ \mu_{\cal{N}}} 
\eeq
The Euler-Lagrange equations can now be applied to \Eq{effEM} to  extract the Maxwell equations (ME)
\beq\label{me}
\partial_\mu \tilde{F}^{\mu\nu} &=& 0  \\ \label{gaussLaw}
\partial_\mu (F^{\mu\nu}-\mu_0 \tilde{M}^{\mu\nu}) &=& \mu_0J^\mu_{\text{free}}~.
\eeq
Below, we first solve the ME in an insulator immersed in a magnetic field with $J^\mu_{\text{free}}=0$, and show that although an electric field is generated, it is too faint to be measured easily.  This is followed by the solutions of the ME in the dipole amplifier, found to be oscillatory with a frequency proportional to the nuclear EDM's. In the case  of a generic ALP $a$, the frequency will be seen to be proportional to $G_{a\gamma\gamma}$, its coupling to photons, times the square root of the ALP number density $n_a$ divided by its mass $m_a$.  In the presence of a CDM ALP background, detection of dark matter {\it hairs}~\cite{Prezeau:2015lxa} (long, dense, filaments of CDM spreading outward from planets) using a dipole amplifier is shown to be feasible.\\[2.25mm]
{\bf EDM: INSULATOR.} We begin by solving ME when $J^\mu_{\text{free}}=0$ to clarify the conditions under which an EDM-induced electric field can be detected in that case.

\begin{figure}
\resizebox{7.5 cm}{!}{\includegraphics*{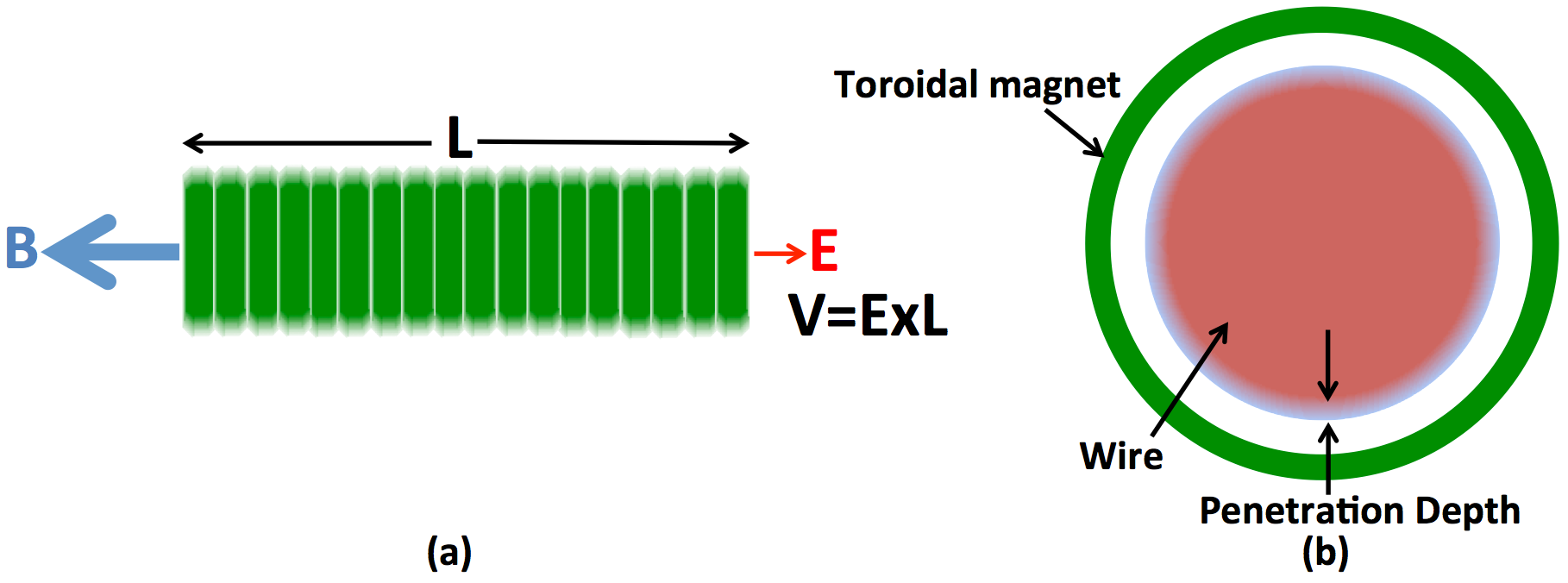}}
\caption{Graph (a) shows a solenoid immersing an insulator in a magnetic field $\vec{B}$.  An electric field $\vec{E}$ is generated aligned with $\vec{B}$.   Graph (b) shows a cross-section of a dipole amplifier: a toroidal magnet immersing a superconducting loop in a magnetic field.  A supercurrent aligned with the magnetic field is generated by the wire material's nuclear EDM, or a background ALP, within the penetration depth $\lambdal$.}\label{amplifier}
\end{figure}

In a bulk material, a nuclear electric dipole moment will be either parallel or anti-parallel to its magnetic moment. If a powerful magnetic field $\vec{B}$ polarizes a substantial fraction of the nuclei in a material, the nuclear EDM is  also amplified thanks to the appearance of a macroscopic electric field $\propto\langle\vec{\sigma}\rangle$, the average magnetic polarization of the bulk material.  Since, $\langle\vec{\sigma}\rangle\propto\vec{H}$, it follows that $\vec{E}\propto\vec{H}$.  Indeed, consider  \Eq{gaussLaw} for $F^{i0}$ 
\beq
\partial^i \left( E^i - c\mu_0 \tilde{M}^{i0} \right) = 0~.
\eeq
The non-relativistic limit of $\tilde{M}^{i0}$ (ignoring ALP treated below)
\beq 
\tilde{M}^{i0}\! \!\!&=&\!\! \!-cP^i\!+\!  c\frac{\mu_\text{s}  d_\text{e}^{\cal{N}}\! \bar{\psi} i\sigma^{i0}\!\gamma_5\psi}{ \mu_{\cal{N}}}  ~~~~~~~~~~~~~~~~~~  \nonumber \\
c\frac{\mu_\text{s}  d_\text{e}^{\cal{N}}\! \bar{\psi} i\sigma^{i0}\!\gamma_5\psi}{ \mu_{\cal{N}}} \! \!&\rightarrow&\! \!-c \mu_{\text{s}}\frac{d_{\text{e}}^{\cal{N}}}{\mu_{\cal{N}}}\langle\sigma^i\rangle_{\cal{N}}   \!=\!  \!-\! c\mu_{\text{s}}\frac{d_{\text{e}}^{\cal{N}}}{\mu_{\cal{N}}}  \chi_{\cal{N}}\!(T)\! H^i
\eeq
where  $\chi_{\cal{N}}(T)$ is the temperature-dependent nuclear magnetic susceptibility of the insulator, $\langle \sigma^i \rangle_{\cal{N}}$ is the average nuclear magnetization, and $H^i$ is the corresponding magnetic field strength.  Since the electric polarizability of insulators is generally small and proportional to the EDM-induced electric field, we can neglect the $\vec{P}$ contribution.  The   ME in the insulator are now
\beq\label{gauss1}
\vec{\nabla}\cdot\vec{B}&=&0 \\ \label{faraday}
\vec{\nabla}\!\times\!\vec{E} + \frac{\partial\vec{B}}{\partial t}&=&0\\ \label{gauss}
\vec{\nabla}\cdot(\vec{E} + \frac{c^2\mu_0}{\mu_{\cal{N}}}d_{\text{e}}^{\cal{N}} \mu_{\text{s}} \chi_{\cal{N}}(T) \vec{H}) &=& 0 \\ \label{ampere}
\vec{\nabla}\!\times\!\vec{H} \! - \epsilon_0 \! \frac{\partial}{\partial t} (\vec{E} \!+\! \frac{c^2\mu_0}{\mu_{\cal{N}}}d_{\text{e}}^{\cal{N}} \mu_{\text{s}} \chi_{\cal{N}}(T) \vec{H}) &=& 0
\eeq
We can write down a solution to the ME from \Eq{gauss}:
\beq\label{gaussSol}
\vec{E} =- \frac{c^2\mu_0}{\mu_{\cal{N}}}d_{\text{e}}^{\cal{N}} \mu_{\text{s}} \chi_{\cal{N}}(T) \vec{H}~.
\eeq
Assuming $\chi_{\cal{N}}(T)$ to be spatially uniform in the insulator, the fields are seen to be static using $\vec{\nabla}\!\times\!\vec{E}\propto\vec{\nabla}\!\times\!\vec{H}=0$ in \Eq{faraday}.  Assuming that  the EDM of the material stems entirely from the nucleus which acts like an ideal paramagnetic, we can  rewrite $ \chi_{\cal{N}}(T) \vec{H}$
\beq\label{brillouin}
\vec{E} \!&=&\! - c^2\mu_0 d_{\text{e}}^{\cal{N}} \mu_{\text{s}} N  B_I(x) \langle\hat{\sigma}\rangle_{\cal{N}} ~,\\ 
B_I(x)\!&\equiv&\! \frac{2I+1}{2I}\text{tanh}\left(\! \frac{2I+1}{2I}x \! \right) \!-\! \frac{1}{2I} \text{tanh}\left( \!\frac{x}{2I}  \!\right)
\eeq
where $x\equiv \mu_{\cal{N}}\vec{B}/(k_\text{B}T)$, $ \langle\hat{\sigma}\rangle_{\cal{N}}$ is the average nuclear magnetization direction, $N$ is the number of nuclei per unit volume and $I$ is the nuclear spin.  To get a sense of the magnitude of that electric field, consider the case of ${}^{127}$I with parameters given in Tab.~\ref{expParam}: $|E|=3.9\cdot10^{-14}$~V/m for $d_{\text{e}}^{\cal{N}}\sim10^{-28}\text{e}\cdot\text{cm}$.  A solenoid coiled around an iodine wire (\Fig{amplifier}a) requires $L=2.5\cdot10^4$~m for a detection on a nanovoltmeter.  For iodine, this may be too conservative because of its relatively large nuclear spin (and, perhaps, the atomic enhancements of the unpaired electron EDM~\cite{Pospelov:2005pr}), but the dipole amplifier described next is far more sensitive and the main topic of this letter.\\[2.25mm]
{\bf EDM: DIPOLE AMPLIFIER.} The dipole amplifier can easily measure a  $d_{\text{e}}^{\cal{N}}\sim10^{-31}\text{e}\cdot\text{cm}$.  It is composed of a superconducting wire inside a toroidal magnet (\Fig{amplifier}b).  The physics behind the amplifier is intuitive: if a bulk material, with a nuclear magnetic moment,  immersed in a magnetic field generates an electric field given by \Eq{gaussSol}, immersing a superconducting loop in a toroidal magnetic field will generate a current, increasing with time, within the penetration depth of that superconducting loop.  Measuring that current will therefore provide a direct measurement of the nuclear EDM of the superconducting material.  Note that the magnetic field magnitude must be less than the critical field of the superconductor, $B_{\text{c}}$.  Nuclear EDM experiments in atoms like vaporized $^{199}$Hg have a relatively low $\bc\sim4.1\cdot10^{-2}$~T.  Since magnetic fields radially  decay exponentially inside a superconducting wire, the weak field approximation of \Eq{brillouin}, Curie's law, will be used to solve ME
\beq\label{curieSol}
\vec{E} =- c^2\mu_0 d_{\text{e}}^{\cal{N}} \mu_{\text{s}} \frac{(I+1)}{I} \frac{N}{3} \frac{\mu_{\cal{N}}\vec{B}}{k_{\text{B}}T}~.
\eeq
with the constraint $\mu_{\cal{N}}\vec{B}\ll k_{\text{B}}T$.  In the case of $^{199}$Hg with a $\mu_{\cal{N}}=0.88\mu_{\text{N}}$  and $|\vec{B}|\sim10^{-2}$~T, this corresponds to the weak constraint $T\gg 10^{-5}$~K.

\begin{figure}[t]
\resizebox{7.5 cm}{!}{\includegraphics*{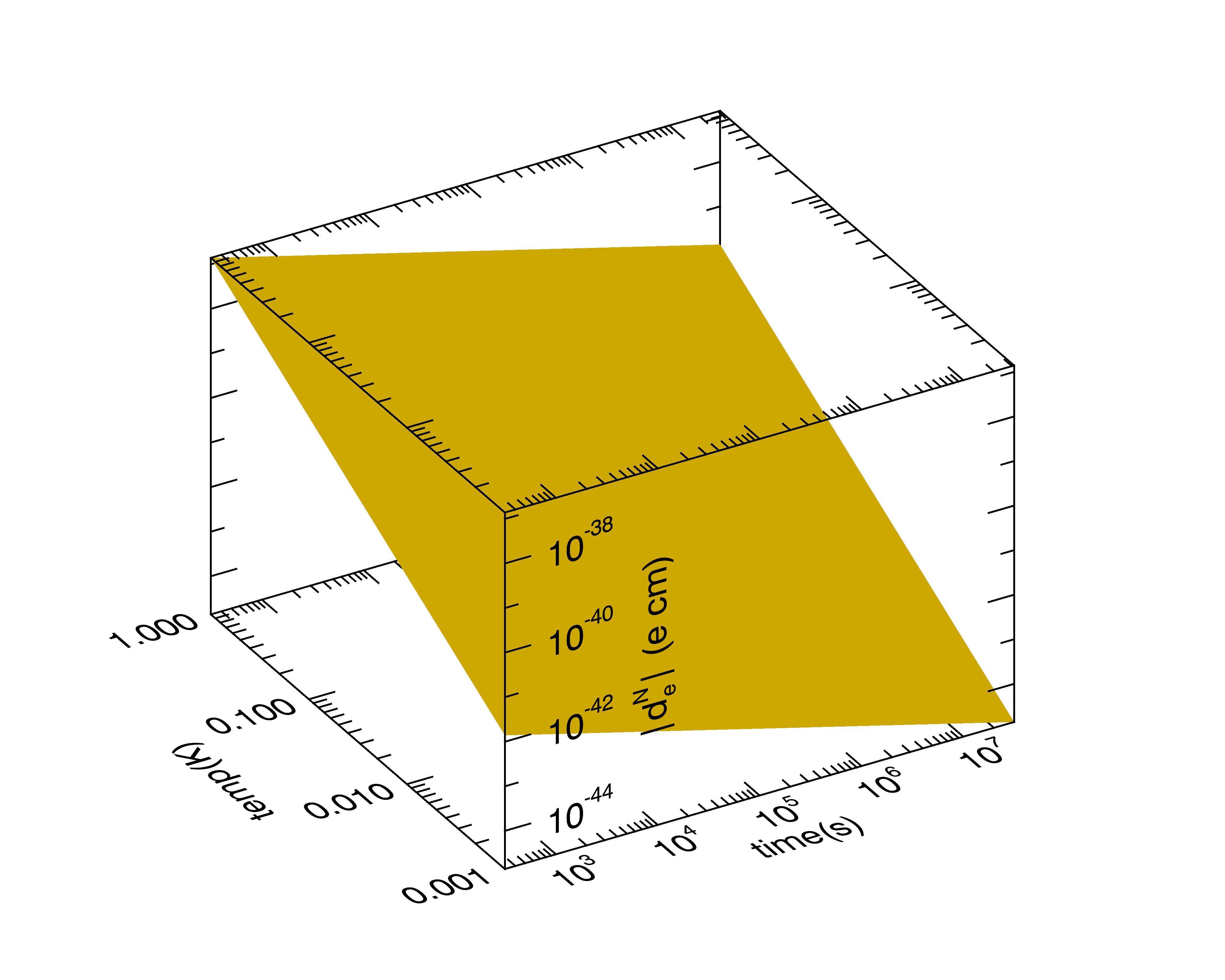}}
\caption{Detectability of $^{199}$Hg nuclear EDM as a function of time and temperature for an experimental setup capable of detecting a 1~pA current.  The parameters used are given in Tab.~\ref{expParam}.  The smallest $|\den|$ detectable after one year at $T=10^{-3}$~K is $|\den|\sim10^{-45}\text{e}\cdot\text{cm}$.  After 315~s, an experiment running at 1~K is already sensitive to $|\den|\sim10^{-37}\text{e}\cdot\text{cm}$.  Current limits on neutron EDM are $\sim3\cdot10^{-26}\text{e}\cdot\text{cm}$ projected to reach $10^{-28}\text{e}\cdot\text{cm}$ at best.}\label{denTimeTemp}
\end{figure}

The ME describing the dipole amplifier are the same as above except for a free current density, $J^\mu_{\text{free}}=(0,\vec{J})$, on the rhs of \Eq{ampere}.   \Eq{curieSol} is the solution of \Eq{gauss} and only \Eq{gauss1}, \Eq{faraday} and \Eq{ampere} (with $\vec{J}$ on the rhs) need to be solved. Let $\vec{B}$ satisfying \Eq{gauss1} be parametrized by $B_l$ and $B_\phi$ 
\beq
\vec{B}=\text{e}^{-(R_{\text{w}}-\rho)/\lambda_{\text{L}}} B_0 ( B_l\hat{l} + B_\phi\hat{\phi} )~,~~\vec{\nabla}\cdot\vec{B}=0~,
\eeq
where $R_{\text{w}}$ is the radius of the wire, $\rho$ is the radial coordinate from the center of the wire, $\hat{l}$ is the wire direction, $\hat{\phi}$ is the azimuthal direction and $\lambdal$ is the London penetration depth. At time $t=0$, we expect $B_l=1$ and $B_\phi=0$.  Neglecting the magnetization $\vec{M}\ll\vec{B}/\mu_0$, the equations to solve and their solutions are
\beq
\vec{\nabla}\!\times\!\vec{E} + \frac{\partial\vec{B}}{\partial t}&=&0\\
\vec{\nabla}\!\times\!\vec{B}  &=& \mu_0\vec{J}  \\
B_l&=& \cos(\omegad t) \\
B_\phi&=& -\sin(\omegad t) \\
\omegad &\equiv& c^2\mu_0 \frac{ \mu_{\text{s}} d_{\text{e}}^{\cal{N}} }{\lambdal}  \frac{(I+1)}{I} \frac{N}{3} \frac{\mu_{\cal{N}}}{k_{\text{B}}T} \\
\vec{J}&=&\frac{  \text{e}^{-(R-\rho)/\lambda_{\text{L}}} }{\lambdal} \frac{B_0}{\mu_0} ( B_\phi\hat{l} - B_l\hat{\phi} )
\eeq
These solutions  satisfy the London Equations.  They also feature a characteristic timescale $|\omegad|^{-1}$ during which the magnetic field and supercurrents in the penetration depth oscillate between their $\hat{l}$ and $\hat{\phi}$ components.  Substituting in parameters for $^{199}$Hg (Tab.~\ref{expParam}) in $\omegad$  assuming a standard model value $d_{\text{e}}^{\cal{N}}=10^{-31}$~e$\cdot$cm, we obtain $T|\omegad|=4.6\cdot10^{-12}$~K$\cdot$Hz.  At $T=1$~K, this corresponds to a period of 7000~years so that an experiment would need to run at $T~\sim10^{-4}$~K to see a full oscillation of the fields and currents in the superconductor.  Although seeing the full oscillation would likely yield the most accurate results, observing a linear rise of a  $\hat{l}$-component supercurrent  for small $\omegad t$  gives fast results and sensitivity to $d_{\text{e}}^{\cal{N}}\ll 10^{-31}$.  For $|\omegad| t\ll1$, $B_{\phi}\cong-\omegad t$ 
\begin{table}
\begin{center}
\begin{tabular}{| c | c |c |c| c| c| c| c   | c |}
\hline
   & $\lambdal$ (\AA) & $\mu_{\cal{N}}$ ($\mu_\text{N}$) & $N$ (m$^{-3}$) & $\mu_\text{s}$ & $I$  & $R_\text{w}$ (cm)  & $B_0$ (T) \\
   \hline
 $^{127}$I & N/A & 3.3 & 2.3$\cdot10^{28}$ & $10^{-4}$ & 5/2  & N/A & $k_{\text{B}} T/\mu_{\cal{N}}$ \\ \hline
 $^{199}$Hg & 520 & 0.88 & 4$\cdot10^{28}$ & $10^{-3}$ & 1/2  & 1 & $10^{-2}$ \\ \hline
  $^{208}$Pb & 370 & 0 &  N/A  & N/A  &  0  & 1 & $10^{-2}$ \\ \hline
\end{tabular}\caption{Bulk and experimental parameters.}\label{expParam}
\end{center}
\end{table}
\beq\label{currentDens}
& &\lim_{\omegad t\ll1} J_l \!=\! - \text{e}^{-(R-\rho)/\lambda_{\text{L}}} \frac{B_0}{\mu_0} \frac{\omegad t}{\lambdal} 
\eeq
for which  the $l$-component of the current density $J_l\propto\lambdal^{-2}\propto n_{\text{e}}$  the charge carrier density.  Integrating \Eq{currentDens} over $\rho$ and $\phi$ to obtain the current magnitude $j_l$, 
\beq\label{amperage}
j_l=\frac{2\pi R_{\text{w}} B_0}{\mu_0} |\omegad| t ~.
\eeq
Consider an experiment sensitive to $j_l\sim1$~pA (e.g. by inserting a Josephson Junction in the loop or measuring the rising $B_\phi$-field) with $^{199}$Hg parameters from Tab.~\ref{expParam}. Putting $j_l=10^{-12}$~A in \Eq{amperage}, \Fig{denTimeTemp} plots $|d_{\text{e}}^{\cal{N}}|$ as a function of time and temperature showing  detectability levels for  $10^{-3}\!<\!T(K)\!<\!1$ within a year.   
\begin{figure}[t]
\resizebox{7.5 cm}{!}{\includegraphics*{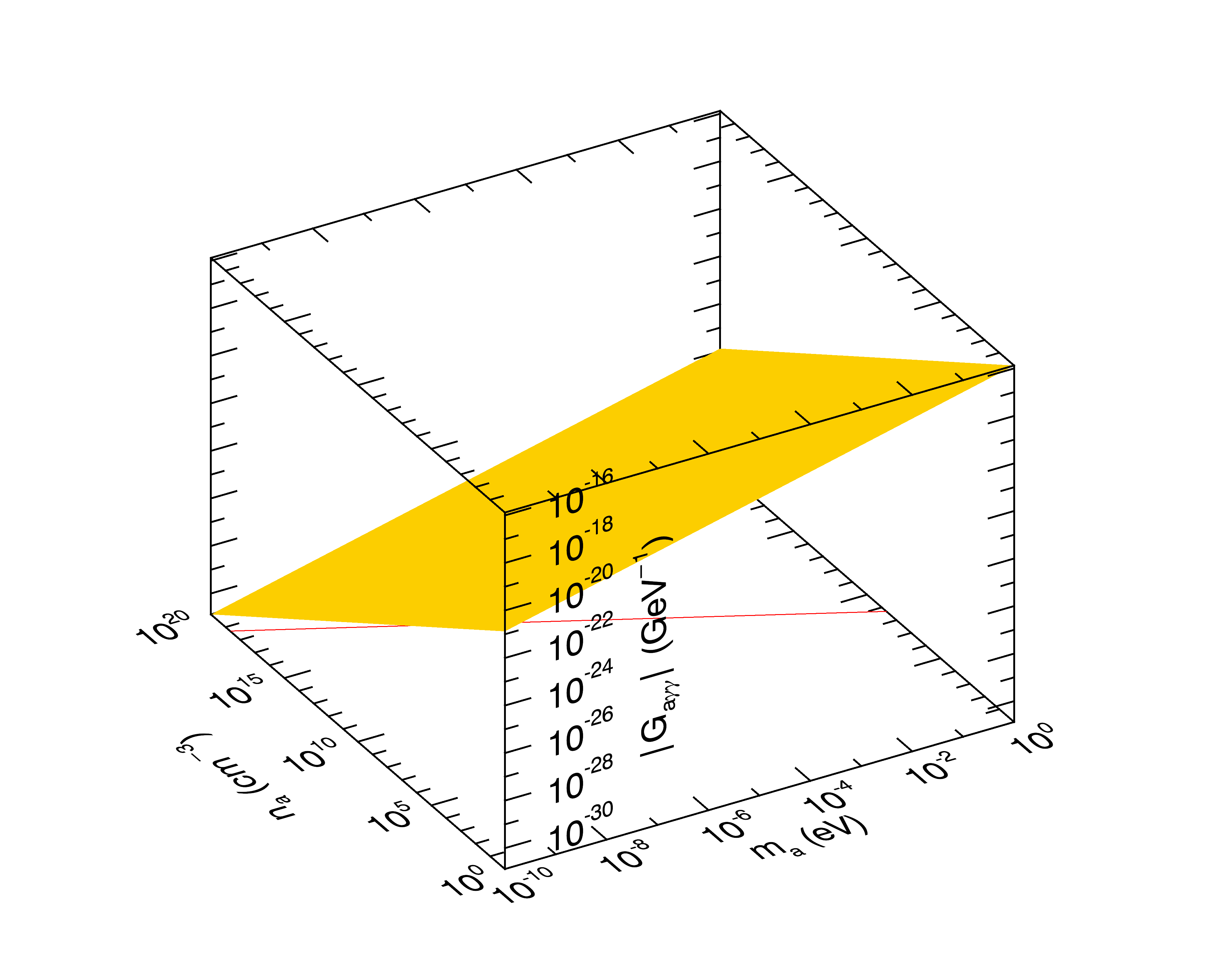}}
\caption{A plot of the CDM ALP parameter space explored by the dipole amplifier sensitive to $j_l\sim1$~pA.  The entire region above the yellow plane would be explored in 60~s which corresponds to all $|G_{a\gamma\gamma}|>10^{-16}$~GeV$^{-1}$ for $m_a<1$~eV and $n_a>1$~cm$^{-3}$, greater than the parameter space considered in Ref.~\cite{Agashe:2014kda}.  The red line corresponds to the local CDM limit~\cite{Agashe:2014kda} $n_am_a=0.45$~GeV/cm$^3$ while the lower-bound from horizontal branch stars ($|G_{a\gamma\gamma}|<6.6\cdot10^{-11}$~GeV$^{-1}$~\cite{Ayala:2014pea}) is much larger.}\label{axionPlot}
\end{figure}
\Fig{denTimeTemp} shows  that a dipole amplifier is sensitive to nuclear EDM down to $|\den|\sim10^{-45}\text{e}\cdot\text{cm}$ at $T=10^{-3}$~K over a year, while a $|\den|\sim10^{-37}\text{e}\cdot\text{cm}$ can be measured in 5~minutes at $T=1$~K. That exceptional sensitivity implies that superconducting materials with smaller  $\mu_\text{s}$ than $^{199}$Hg can be used if the nuclear matrix elements are more easily calculable than the soft nucleus of $^{199}$Hg.\\[1.5mm]
{\bf AXION-LIKE PARTICLES.} Observational and laboratory constraints on the axion have typically relied on detecting or constraining on-shell photons generated by the Primakoff effect.  As the  amplifier relies on measuring a DC electric field induced by a background ALP field, the final photon 4-momentum does not generally satisfy $q^2=0$  expanding the final photon phase space by orders of magnitude: combined with the sensitivity of superconductors to infinitesimal electric fields, this explains why the dipole amplifier is so sensitive compared to other approaches.  For a CDM ALP background, $\omegad$ is given by
\beq\label{omegadalp}
\omega_{\text{D}} &=&  \frac{|G^i_{a\gamma\gamma}| a_i}{2  } \frac{c}{\lambda_{\text{L}}(T)}~.
\eeq
The $a_i$ are constrained by the covariant normalization condition $\int a_{i}^2 2m_{a_i}\text{d}V=n_{i}$ where there are $n_{i}$ CDM ALP of type $i$ in volume $V$.  For uniform ALP densities, the total number $N$ of CDM ALP in volume $V$ is
\beq
\sum_{i} 2m_{a_i} a_{i}^2 = \frac{N}{V}\equiv n_a \xrightarrow{\text{single species}} a=\sqrt{\frac{n_a}{2m_a}}~.
\eeq
Consider an experiment that runs for 1 minute with a dipole amplifier sensitive to 1~pA currents, and a CDM ALP background composed of a generic species  $a$.  To suppress a signal from nuclear EDM, choose  a diamagnetic superconducting material with a null nuclear spin like $^{208}$Pb. Detecting a $j_l>1$~pA in 1 minute requires
\beq
|G_{a\gamma\gamma}|\sqrt{ \frac{n_a}{2m_a} } \! >\! \frac{2\lambdal}{c\text{60s}} \frac{(10^{-12}\text{A}) \mu_0}{2\pi R_\text{w}B_0}~. 
\eeq
For $^{208}$Pb and Tab.~\ref{expParam} values, the sensitivity of the dipole amplifier is plotted in \Fig{axionPlot} where it is seen that within a minute, the entire parameter space $|G_{a\gamma\gamma}|>10^{-16}$~GeV$^{-1}$ for $m_a<1$~eV and $n_a>1$~cm$^{-3}$ is explored.  Some of that parameter space is already constrained by astrophysical sources such as limits stemming from horizontal branch stars ($|G_{a\gamma\gamma}|<6.6\cdot10^{-11}$~GeV$^{-1}$)~\cite{Ayala:2014pea} and a local CDM density of $n_am_a=0.45$~GeV/cm$^3$ assuming all CDM to be ALP.

We will consider the case of relativistic ALP, and methods to disentangle their signal in a dipole amplifier from CDM ALP, in a future paper.\\[2.25mm]
{\bf DENSE DARK MATTER HAIRS.} The dipole amplifier could also be used to detect dark matter hairs~\cite{Prezeau:2015lxa} if CDM ALP are found to exist: assuming a probe  passes through a 1~m wide hair in $10^{-4}$~s, a detectable  pA current spike could be detected as long as 
\beq
|G_{a\gamma\gamma}|\sqrt{ \frac{n_h}{(2m_a)} } > 4.9\cdot10^{-27}
\eeq
where $n_h$ is the hair ALP number density.  As a reference point, consider typical values often used for axions~\cite{Agashe:2014kda}: if $n_a\sim10^{13}$/cm$^3$, $m_a\sim10^{-5}$~eV and $G_{a\gamma\gamma}\sim10^{-12}$/GeV, $|G_{a\gamma\gamma}|\sqrt{ n_a/(2m_a) }=6.2\cdot10^{-20}$.  As discussed in ~\cite{Prezeau:2015lxa}, hairs are potential sources of unique data sets of the  fine-grained dark matter streams criss-crossing the solar system, including their velocity and density distributions.  This data, unobtainable any other way, is highly relevant for structure formation and cosmology and would act as a constraint on simulations of halo formation~\cite{Vogelsberger:2010gd}.  In addition, hairs  contain radial density tomographic data of the planets and moons from which they spread out, providing information that could be used to better understand planetary formation and the history of the solar system.

\begin{acknowledgments}

The author is grateful to Brad Plaster and Takeyasu Ito for useful comments and suggestions.  The research was carried out at the Jet Propulsion Laboratory, California Institute of Technology, under a contract with the National Aeronautics and Space Administration. \copyright 2016 California Institute of Technology. Government sponsorship acknowledged.

\end{acknowledgments}

\end{document}